\documentclass[a4paper, 10pt, conference]{ieeeconf}      

\IEEEoverridecommandlockouts                              

\usepackage{graphics} 
\usepackage{epsfig} 
\usepackage{amsmath} 
\usepackage{amssymb}  
\usepackage{color,soul}
\usepackage{balance}

\usepackage{acronym}

\acrodef{MMSE}{minimum mean square error}
\acrodef{KF}{Kalman filter}
\acrodef{i.i.d.}{independent and identically distributed}
\acrodef{QCD}{Quickest Change Detection}
\acrodef{LFD}{Least Favorable Distribution}

\title{\LARGE \bf
Bayesian Quickest Change Detection of an Intruder in Acknowledgments for Private Remote State Estimation
}

\author{Justin M. Kennedy, Jason J. Ford, and Daniel E. Quevedo
\thanks{$^{1}$ J. M. Kennedy, J. J. Ford, and D. E. Quevedo are with the School of Electrical Engineering and Robotics, Queensland University of Technology, 2 George St, Brisbane QLD, 4000 Australia. The authors acknowledge continued support from the Queensland University of Technology (QUT) through the Centre for Robotics.
        {\tt\small \{j12.kennedy, j2.ford, daniel.quevedo\}@qut.edu.au}}%
}

\usepackage{eso-pic}
\usepackage{url}
\AddToShipoutPicture*{%
     \AtTextUpperLeft{%
         \put(-3.5,10){
           \begin{minipage}{\textwidth}
              \scriptsize
              \MakeUppercase{Preprint version of an Australian and New Zealand Control Conference article;\\ final version available at} \url{https://doi.org/10.1109/ANZCC56036.2022.9966866}
           \end{minipage}}%
     }%
}

\begin{document}

\maketitle
\thispagestyle{empty}
\pagestyle{empty}

\begin{abstract}
For geographically separated cyber-physical systems, state estimation at a remote monitoring or control site is important to ensure stability and reliability of the system.
Often for safety or commercial reasons it is necessary to ensure confidentiality of the process state and control information.
A current topic of interest is the private transmission of confidential state information.
Many transmission encoding schemes rely on acknowledgments, which may be susceptible to interference from an adversary.
We consider a stealthy intruder that selectively blocks acknowledgments allowing an eavesdropper to obtain a reliable state estimate defeating an encoding scheme.
We utilize Bayesian Quickest Change Detection techniques to quickly detect online the presence of an intruder at both the remote transmitter and receiver.
\end{abstract}

\section{INTRODUCTION}

Security of cyber-physical systems is a current topic of interest \cite{Ishii2022SecurityResilienceControl} with three key problems: ensuring confidentiality, integrity, and availability of data over a network.
One class of problems in systems and control design, is the state estimation of a remote process over an unreliable wireless network in the presence of an eavesdropper.

Designs to ensure privacy against eavesdropping have been addressed in several works including
transmission encoding schemes \cite{Lucke2022IEEETransactionsonAutomaticControlCodingsecrecyremote} \cite{Tsiamis2020IEEETransactionsonAutomaticControlStateSecrecyCodes} 
and scheduling of transmissions \cite{Leong2019IEEETransactionsonAutomaticControlTransmissionSchedulingRemote}.
These information privacy techniques rely on acknowledgment of successful packet receipt by the legitimate remote estimator to the transmitter, such that the transmitter and legitimate receiver are able to remain in sync.

Conversely, to improve an eavesdropper's performance, adversarial designs have exploited the vulnerability in packet acknowledgments, including
hacking the scheduling policy  \cite{Lu2021IEEETransactionsonAutomaticControlStealthyhackingsecrecy} and
fake acknowledgment transmission \cite{Ding2019IEEETransactionsonControlofNetworkSystemsDoSAttacksRemote}.
A powerful eavesdropper was considered in \cite{Ding2021IEEETransactionsonAutomaticControlRemoteStateEstimation} that could perform both eavesdropping and acknowledgment blocking tasks simultaneously, with an optimal acknowledgment attack strategy on a transmission scheduling policy.
To degrade the legitimate receiver's performance, \cite{Cheng2020IEEETransactionsonAutomaticControlEventBasedStealthy} proposed an event based attack in the acknowledgments, while \cite{Zhang2018IEEETransactionsonControlofNetworkSystemsDoSAttackEnergy} considered the transmission energy usage required to block acknowledgments, devising a strategy to balance performance degradation with limited energy usage.
Techniques to combat integrity and denial of service attacks have included distributed estimation \cite{Yang2022AutomaticaAdaptivedistributedKalman} with application to power networks,
and watermarking of transmissions to enable detection of data modification \cite{Naha2022IEEETransactionsonAutomaticControlSequentialdetectionReplay}.

In this paper, we consider \cite{Tsiamis2020IEEETransactionsonAutomaticControlStateSecrecyCodes} which proposed a transmission encoding scheme using innovations.
When an eavesdropper misses a packet that the legitimate receiver obtains, the eavesdropper is unable to recover, and its estimation error will constantly grow\footnote{The eavesdropper's estimation error will grow to infinity in the case of unstable dynamics \cite{Tsiamis2020IEEETransactionsonAutomaticControlStateSecrecyCodes} or to the open loop estimation error in the case of stable dynamics \cite{Tsiamis2018StateSecrecyCodes}.}.
We propose a powerful eavesdropper, in the sense of \cite{Ding2021IEEETransactionsonAutomaticControlRemoteStateEstimation}, which to obtain a reliable estimate, blocks the receipt acknowledgments.
A simple method for the attacker, but trivial to detect, would be to block all acknowledgments.
We propose a stealthy acknowledgment blocking method that selectively blocks based on the eavesdropper's estimation performance.
While the approach may appear stealthy, we outline an intruder detection strategy using statistical analysis of the age of information available to the transmitting sensor as well as to the legitimate receiver.

In Section~\ref{sec:remotestateest}, we outline the private remote state estimation technique of \cite{Tsiamis2020IEEETransactionsonAutomaticControlStateSecrecyCodes} and formulate the intruder detection problem.
We present Bayesian Quickest Change Detection stopping rules to detect an intruder in Section~\ref{sec:intruderdetection}.
We illustrate the performance of the detectors in Section~\ref{sec:simulation}, and conclude in Section~\ref{sec:conclusion}.

\section{REMOTE STATE ESTIMATION}
\label{sec:remotestateest}

We now formalize our remote state estimation scenario over an unreliable wireless network using the transmission encoding technique of \cite{Tsiamis2020IEEETransactionsonAutomaticControlStateSecrecyCodes}.
We then propose the attack method in the acknowledgment channel.

\subsection{System Dynamics \& Sensor Estimate}
Consider the following unstable discrete-time LTI process with state $x_k \in \mathbb{R}^n$ and measurement $y_k \in \mathbb{R}^m$
\begin{equation*}
    x_{k+1} = A x_k + w_k,  \quad y_k = C x_k + r_k
\end{equation*}
where $w_k$ and $r_k$ are zero-mean Gaussian distributed process and measurement noise with covariance $Q$ and $R$, respectively, and $A$ has eigenvalues outside the unit circle.
The encoding scheme in \cite{Tsiamis2020IEEETransactionsonAutomaticControlStateSecrecyCodes} utilizes unstable dynamics in the encoding to let the covariance of the eavesdropper's estimation error diverge to infinity.
We assume that the pair $(A,C)$ is observable and that the pair $(A,\sqrt{Q})$ is controllable.
These standard dynamics encode many processes considered in control design.

We assume that the initial state of the process $x_0$ is a Gaussian random variable with zero mean and covariance $\Sigma_0$, and that the two covariances $Q$ and $\Sigma_0$ are positive definite.
Additionally, we consider that $w_k$, $r_k$, and $x_0$ are uncorrelated.
Finally, we assume that $A$, $Q$, $R$, and $\Sigma_0$ are public, available to all network users, but the realizations of the initial state $x_0$ and noise $w_k$ are not known.

By operating a Kalman filter at the sensor, an optimal state estimate $\hat{x}_k^s = E[x_k | y_0, \dots, y_k]$
with estimation error covariance $P_k^s = E[(x_k - \hat{x}_k^s)(x_k - \hat{x}_k^s)^\mathsf{T} | y_0, \dots, y_k]$ can be produced.
This is the best state estimate given access to all available measurements $y_k$ \cite{Anderson1979OptimalFiltering}.

\subsection{Network Model}
We consider that a remote operator requires a reliable estimate of the system state for the purposes of remote monitoring or control.
To obtain this estimate, the sensor transmits state estimate information over a wireless network to the remote operator.
However, the transmitted packets can also be received by an eavesdropper.

To ensure privacy from eavesdroppers, the sensor cleverly encodes the current state estimate $\hat{x}_k^s$ into a packet $z_k$ which can be decoded by the legitimate estimator but not always by an eavesdropper\footnote{We will describe how $z_k$ is formed in Section~\ref{subsec:encoding}.}.
The sensor transmits this encoded packet $z_k$ over an unreliable wireless channel to the intended legitimate estimator.
Upon receipt of the packet, the legitimate estimator decodes the packet, and acknowledges receipt over a separate unreliable acknowledgment channel.

\begin{figure}
  \centering
  \resizebox{0.45\textwidth}{!}{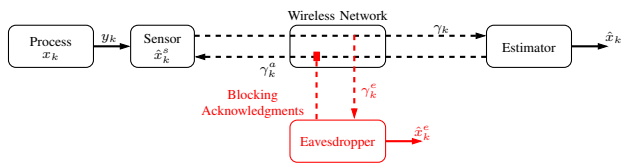}
  \caption{Architecture of the channel environment. A remote process is monitored by a sensor, which sends an encoded state update over an unreliable network that can be received by the legitimate estimator and an eavesdropper. The packet is encoded with respect to the last received acknowledged to ensure privacy. The eavesdropper intrudes, blocking the acknowledgments to ensure that it has a reliable state estimate.}
  \label{fig:networkmodel}
\end{figure}

In this work, we consider a powerful intruder in the network.
Our intruder eavesdrops on the transmissions receiving packets and also interferes blocking acknowledgments to control the encoding scheme.
We outline our attack technique below in reference to the encoding scheme.
This network model is visualized in Figure~\ref{fig:networkmodel}.

Let us define $\gamma_k \in \{0,1\}$ as an indicator variable denoting successful reception at the legitimate estimator where
\begin{equation*}
    \gamma_k = \begin{cases} 1, \mbox{ if the packet is received,} \\ 0, \mbox{ if a packet dropout occurs,} \end{cases}
\end{equation*}
and we similarly define for $\gamma_k^e \in \{0,1\}$ at the eavesdropper, and $\gamma_k^a \in \{0,1\}$ at the sensor for acknowledgments.
We assume that the three channel outcomes 
are \ac{i.i.d.}, and independent to the initial state of the process and the process noise.
We note that acknowledgments are only sent by the legitimate estimator on successful receipt, and are then dependent on the legitimate estimator channel.
However, each acknowledgment transmission over the acknowledgment channel is an independent event, and we thus consider the acknowledgment channel outcomes as conditionally independent and identically distributed. 

Let us define the three channel qualities as Bernoulli random variables where for the legitimate estimator $P(\gamma_k = 1) = \alpha$, for the eavesdropper $P(\gamma_k^e = 1) = \alpha_e$, and for acknowledgment $P(\gamma_k^a = 1) = \alpha_a$, where $0 \leq \alpha, \alpha_e, \alpha_a \leq 1$.
We consider that the legitimate estimator's and sensor acknowledgment channel quality $\alpha$ and $\alpha_a$ are known to both the legitimate estimator and sensor.
The eavesdropper's channel quality $\alpha_e$ may be unknown.

We define the measurement at the legitimate estimator as $\hat{y}_k = \gamma_k z_k$ and eavesdropper as $\hat{y}_k^e = \gamma_k^e z_k$.
The information available to the legitimate estimator at time $k$ is the set of measurements $\mathcal{I}_k = \{\gamma_0, \hat{y}_0, \dots, \gamma_k, \hat{y}_k\}$ and for the eavesdropper $\mathcal{I}_k^e = \{\gamma_0^e, \hat{y}_0^e, \dots, \gamma_k^e, \hat{y}_k^e\}$.

We define the state estimate and estimation error covariance at the legitimate receiver as
\begin{equation*}
    \hat{x}_k = E[x_k | \mathcal{I}_k] \quad P_k = E[(x_k - \hat{x}_k)(x_k - \hat{x}_k)^\mathsf{T} | \mathcal{I}_k]
\end{equation*}
and for the eavesdropper as
\begin{equation*}
    \hat{x}_k^e = E[x_k | \mathcal{I}_k^e] \quad P_k^e = E[(x_k - \hat{x}_k^e)(x_k - \hat{x}_k^e)^\mathsf{T} | \mathcal{I}_k^e] .
\end{equation*}

\subsection{Transmission Encoding and Decoding}
\label{subsec:encoding}
The goal of the encoding scheme is to minimize the legitimate estimator's estimation error covariance $P_k$ and maximize the eavesdropper's estimation error covariance $P_k^e$.
Tsiamis et al. \cite{Tsiamis2020IEEETransactionsonAutomaticControlStateSecrecyCodes} proposed to transmit an innovation, the difference between the current state estimate and the last successfully acknowledged packet propagated through the dynamics
\begin{equation}
	    z_k = \hat{x}_k^s - A^{k-t_k} \hat{x}_{t_k}^s
	    \label{eq:thepacket}
\end{equation}
where $t_k$ is the time that a packet was last successfully received by the legitimate estimator $(\gamma_{t_k} = 1)$ and the acknowledgment was successfully received $(\gamma_{t_k}^a = 1)$.

To decode the packet, the legitimate receiver adds the measurement to the known state propagated from time $t_k$
\begin{equation*}
    \hat{x}_k = \hat{y}_k + A^{k-t_k} \hat{x}_{t_k} .
\end{equation*}
In the case of the successful receipt $(\gamma_k = 1)$ then
\begin{equation*}
    \hat{x}_k = z_k + A^{k-t_k} \hat{x}_{t_k} = \hat{x}_k^s - A^{k-t_k} \hat{x}_{t_k}^s + A^{k-t_k} \hat{x}_{t_k} = \hat{x}_k^s
\end{equation*}
noting that as the previous packet $z_{t_k}$ was successfully received then $\hat{x}_{t_k} = \hat{x}_{t_k}^s$.
The estimation error covariance is then $P_k = P_k^s$, the same estimation error covariance as at the sensor.
In the case of a dropout $\gamma_k = 0$, the dynamics are utilized to produce a state prediction,
    $\hat{x}_k = A \hat{x}_{k-1}$,
with growing state estimation error covariance $P_k = A P_{k-1} A^\mathsf{T} + Q$ from the last received packet\footnote{Due to acknowledgment dropouts, this may be more recent than $t_k$. }. 

As the acknowledgment channel is unreliable, the legitimate receiver then requires knowledge of the last acknowledged time $t_k$ to remain in sync.
The packet \eqref{eq:thepacket} is modified to also include the time $t_k$
\begin{equation*}
	z_k = \{\hat{x}_k^s - A^{k-t_k} \hat{x}_{t_k}^s, t_k\} .
\end{equation*}

It is shown in \cite{Tsiamis2020IEEETransactionsonAutomaticControlStateSecrecyCodes} that utilizing this encoding scheme, the state remains secret when a passive eavesdropper misses a packet.
We briefly outline how a single missed packet results in eavesdropper's state estimate degrading.
Consider that an eavesdropper has been in sync up to time $t$, such that $\hat{x}_{t}^e = \hat{x}_{t}^s$.
At some time $\ell > t$, the critical event occurs where the eavesdropper drops the packet $\gamma_{\ell}^e = 0$ which the legitimate estimator successfully receives $\gamma_\ell = 1$ and acknowledges $\gamma_\ell^a = 1$.
Under the assumption of unreliable and independent channel outcomes, and that the ``critical'' event has non-zero probability $P(\gamma_{\ell}^e = 0, \gamma_\ell = 1, \gamma_\ell^a = 1) > 0$, the critical event will occur infinitely often \cite[Remark 2]{Tsiamis2020IEEETransactionsonAutomaticControlStateSecrecyCodes}.
As $z_{\ell}$ was not received, the eavesdropper uses the dynamics to produce a prediction
    $\hat{x}_{\ell}^e = A^{\ell - t} \hat{x}_{t}^e$.
Note that $\hat{x}_{\ell}^e \neq \hat{x}_{\ell}^s$.
The next encoded packets $z_k$ for $k>\ell$ will now use the last acknowledged time as $\ell$.
Consider a later time $\kappa > \ell$ where the packet is $z_\kappa = \{ \hat{x}_\kappa^s - A^{\kappa - \ell}\hat{x}_\ell^s, \ell \}$, and
that the eavesdropper successfully receives this packet $\gamma_\kappa^e= 1$.
The eavesdropper's state estimate is then
\begin{align*}
    \hat{x}_\kappa^e &= z_\kappa + A^{\kappa-\ell} \hat{x}_{\ell}^e = \hat{x}_\kappa^s - A^{\kappa-\ell} \hat{x}_{\ell}^s + A^{\kappa-\ell} \hat{x}_{\ell}^e
\end{align*}
The error between $\hat{x}_{\ell}^e$ and $\hat{x}_{\ell}^s$ is magnified through the unstable dynamics $A^{\kappa-\ell}$,
such that the error between the state estimate and the true state will grow: $P_k^e \rightarrow \infty$ as $k \rightarrow \infty$.
See \cite{Tsiamis2020IEEETransactionsonAutomaticControlStateSecrecyCodes} for the detailed proof.

\subsection{Attack Model}
\label{subsec:remotestate:ackintrude}
To remain in sync with the legitimate estimator under the above encoding scheme, an eavesdropper needs to receive all transmitted packets, receive exactly the same packets as the legitimate estimator, or control the instances where the reference time $t_k$ is updated.
If an eavesdropper is powerful enough with a perfect channel $\alpha_e = 1$ or has hacked the legitimate estimator to receive the same packets such that $\gamma_k^e = \gamma_k$ for all $k>0$, then alternative methods to ensure privacy should be considered.

Let us consider that the attacker intrudes in the acknowledgment channel, blocking acknowledgments thereby controlling when $t_k$ is updated.
In the presence of acknowledgment blocking, the encoding scheme of \cite{Tsiamis2020IEEETransactionsonAutomaticControlStateSecrecyCodes} continues to function as the last received and acknowledged time, $t_k$, is shared in the packet.

The most simple method is to block all acknowledgments, such that $t_k$ never updates.
However, it would be trivial for both the sensor and legitimate estimator to detect the presence of such an intruder.
The sensor would receive no acknowledgments, and the receiver would observe that the transmitted $t_k$ never changes.

Let us consider a more stealthy approach, where the attacker makes a decision to block based on its own information.
The attacker only blocks the acknowledgment during the critical event, where the packet was received by the legitimate estimator ($\gamma_k=1$) but not the eavesdropper ($\gamma_k^e = 0$).
In the case that both receive a packet ($\gamma_k=\gamma_k^e=1$), the attacker allows the acknowledgment to pass.
The legitimate estimator will not send an acknowledgment when it has not received a packet, so no intruder action is required.
Importantly, the eavesdropper only acts when the legitimate estimator obtains data and the eavesdropper does not receive data.

\section{INTRUDER DETECTION}
\label{sec:intruderdetection}

Detecting the intrusion becomes more difficult as the sensor is still receiving acknowledgments, and the last acknowledged time $t_k$ is regularly changing.
However, after the intrusion, there will be a statistical change resulting from the eavesdropper's actions in the rate of acknowledgment receipts and in updates of the last acknowledged time $t_k$.

Our problem is to quickly detect the presence of the intruder with the information available at the legitimate estimator $\mathcal{I}_k$.
We present Bayesian \ac{QCD} statistical test and an efficient computational method.
We also pose an alternative detection problem with the acknowledgment information available at the sensor $\mathcal{I}_k^a \triangleq \{\gamma_0^a, \dots, \gamma_k^a \}$.

\subsection{Intruder Detection as QCD at the Receiver}
\label{subsec:qcd}
Let $D_m = k$ denote the time the legitimate estimator successfully receives, $\gamma_k=1$, its $m$th packet where $m \ge 1$.
The packet receipt index $m$ is a non-uniform sub-sampling of the process time $k$.
We define the age of innovation $\mathcal{A}_m$, as the difference between the current time of packet receipt $D_m$ and last successful acknowledged time $t_k$ included in the packet $z_k$
\begin{equation}
    \mathcal{A}_m \triangleq D_m - t_k .
    \label{eq:ageofinfo}
\end{equation}
At some successful packet receipt $m=\lambda$ where $\lambda \ge  1$ the eavesdropper intrudes into the acknowledgment channel, blocking successful acknowledgments.

Before the intrusion, at each time instance $k$, the success of a packet receipt with successful acknowledgment is binomial distributed with probability $P(\gamma_k=1,\gamma_k^a=1) = P(\gamma_k=1) P(\gamma_k^a=1) = \alpha\alpha_a$.
The number of failures in sequence, or the age $\mathcal{A}_m$ is geometrically distributed with parameter $\rho_1=\alpha \alpha_a$ as follows
\begin{equation}
    b_1(\mathcal{A}_m) \triangleq P(\mathcal{A}_m=\kappa)=\rho_1 (1 - \rho_1)^{\kappa-1}
    \label{eq:b1}
\end{equation}
for $\kappa\ge 1$ and 0 otherwise.

After the intruder enters, acknowledgments are blocked when the eavesdropper does not receive the packet.
At each time instance $k$, the success of a packet receipt with successful acknowledgment time is binomial with probability $P(\gamma_k=1,\gamma_k^a=1,\gamma_k^e=1) = P(\gamma_k=1) P(\gamma_k^a=1) P(\gamma_k^e=1) = \alpha \alpha_a \alpha_e$.
The age $\mathcal{A}_m$ is geometrically distributed with parameter $\rho_2=\alpha \alpha_a \alpha_e$ as follows
\begin{equation}
    b_2(\mathcal{A}_m) \triangleq P(\mathcal{A}_m=\kappa)=\rho_2 (1 - \rho_2)^{\kappa-1}
    \label{eq:b2}
\end{equation}
for $\kappa\ge 1$ and 0 otherwise.
The probability of a successful event has decreased by the quality of the eavesdropper's channel.
The age of the information under the intruder scenario will then be probabilistically larger than under the no-intruder scenario.

In this work, we model the commencement of intrusion instance $\lambda \ge 1$ as having geometric distribution $\pi \triangleq \{\pi_m:m\geq 1\}$, 
where
\begin{equation}
    \pi_m\triangleq P(\lambda = m) = \rho_i (1 - \rho_i)^{m-1}
    \label{eq:changegeoprior}
\end{equation}
where $0 < \rho_i < 1$ is a given parameter.

This formulation is amenable in the framework of a standard \ac{QCD} problem, where we look for a change in the statistical properties of an observed process.
Under this change description, the given intrusion time $\lambda \ge 1$, the joint probability density function of the observed packet ages until the $m \ge 1$ packet is given by
\begin{equation*}
    p_\lambda(\mathcal{A}_{[1,m]}) = \Pi_{\ell=1}^{\lambda-1} b_1(\mathcal{A}_\ell) \Pi_{\ell=\lambda}^{m} b_2(\mathcal{A}_\ell) \mbox{ for }\; m \ge 1
\end{equation*}
and we define $\Pi_{\ell=\kappa}^{\nu} b_i(\mathcal{A}_\ell) = 1$ for $\kappa >  \nu$ and $i \in \{1,2\}$ where $b_i$ is defined in \eqref{eq:b1}--\eqref{eq:b2}.

Let $\mathcal{F}_\ell$ denote the filtration generated by $\mathcal{A}_\ell$ for $\ell \geq 1$.
We consider the probability space $(\Omega, \mathcal{F}, P_\lambda)$ where $\Omega$ is the sample space of infinite sequences $\{\mathcal{A}_\ell : \ell \geq 1\}$ and $\mathcal{F} \triangleq \cup_{\ell=0}^\infty \mathcal{F}_\ell$ with the convention that $\mathcal{F}_0 \triangleq \{0,\Omega\}$, and $P_\lambda$ is the probability measure constructed from the joint probability density 
$p_\lambda(\mathcal{A}_{[1,\ell]})$ using Kolmogorov’s extension theorem
\cite{elliott2008hidden}.

Let us now construct a new  probability measure $P_\pi$ from $P_\lambda$ for $\lambda\ge 1$ by averaging in the sense that 
\begin{equation*}
    P_\pi (\mathcal{B}) \triangleq \sum_{m=1}^\infty   \pi_m P_m  (\mathcal{B})
\end{equation*}
for all $\mathcal{B} \in \mathcal{F}$.
We will denote the expectation operator associated with $P_\pi $ as $E_\pi[\cdot]$.

Our goal is to quickly detect the presence of the intruder by designing a stopping time $\tau \ge 1$ that minimizes the following cost
which is the unconstrained (Bayes risk) optimization problem of \cite{Shiryaev1963TheoryofProbability&amp$mathsemicolon$ItsApplicationsOptimumMethodsQuickest}
\begin{equation}
    J(\tau) \triangleq c E_\pi \left[ (\tau - \lambda)^+ \right] + P_\pi(\tau < \lambda),
    \label{eq:bayesrisk}
\end{equation}
where $(\tau - \lambda)^+ \triangleq \max(0,\tau-\lambda)$, and $c$ is the penalty at each time instance before declaring an alert at $\tau$.
The cost $J(\tau)$ balances the impact of expected detection delay against the probability of false alarms.

We let $P_\pi(m < \lambda | \mathcal{F}_m)$ be the no-change posterior probability (we show in Section~\ref{subsec:efficientcalc} how to calculate efficiently).
Following \cite{Unnikrishnan2011IEEETransInfoTheoryMinimaxRobustQuickest}, the optimal stopping rule for 
the problem \eqref{eq:bayesrisk} is
\begin{equation}
    \tau^*=\inf\{m \ge 1: P_\pi(m < \lambda | \mathcal{F}_m) \leq h\} 
    \label{eq:optstoptime}
\end{equation}
for some $h\in (0,1)$ selected to manage the probability of false alarms.

At each successful receipt $\gamma_k = 1$, we compute the no-change posterior $P_\pi(m < \lambda | \mathcal{F}_m)$ using the latest age of information \eqref{eq:ageofinfo}.
We then achieve an optimal stopping time under \eqref{eq:optstoptime} and declare that an intruder has interfered in the acknowledgment channel.

\subsection{Efficient Calculation of the No-Change Posterior}
\label{subsec:efficientcalc}
We now present an efficient recursive method to compute the no-change posterior $P_\pi(m < \lambda | \mathcal{F}_m)$.
Let us define $\hat{Z}_m^1 \triangleq P_\pi(m < \lambda | \mathcal{F}_m)$, and recall $b_1(\cdot)$ and $b_2(\cdot)$ from \eqref{eq:b1}--\eqref{eq:b2}.
Following \cite[Lemma 1]{Ford2020AutomaticainformativenessmeasurementsShiryaevs} for $m \ge 1$, given a sequence of measurements $\mathcal{A}_{[1,m]}$ the no change posterior probability $\hat{Z}_m^1$ is given by the scalar recursion
\begin{equation*}
    \hat{Z}_m^1 = N_m (1 - \rho_i) b_1(\mathcal{A}_m) \hat{Z}_{m-1}^1
\end{equation*}
with $\hat{Z}_0^1 = 1$ and the normalization factor
\begin{equation*}
    N_m^{-1} = b_2(\mathcal{A}_m) + (1-\rho_i)(b_1(\mathcal{A}_m) - b_2(\mathcal{A}_m)) \hat{Z}_{m-1}^1 .
\end{equation*}

\subsection{Alternative Detection at the Sensor}
It is also possible to construct a detector at the transmitting sensor.
Let $D_n^s = k$ denote the time the sensor successfully receives, $\gamma_k^a=1$, its $n$th acknowledgment where $n \ge 1$.
The acknowledgment receipt index $n$ is a non-uniform sub-sampling of the process time $k$, and further sub-sampled to the packet receipt index $m$ above.
Let us define the age of acknowledgment as the difference between the current time of acknowledgment receipt $D_n^s$ and last successful acknowledged time $t_k$ used in the packets $z_k$
\begin{equation}
    \mathcal{A}_n^s \triangleq D_n^s - t_k
    \label{eq:ageofack}
\end{equation}
which we observe is similar to \eqref{eq:ageofinfo}.
At some successful acknowledgment receipt $n=\lambda$ where $\lambda \ge 1$ the eavesdropper intrudes into the acknowledgment channel, blocking successful acknowledgments.
The age of acknowledgment can be described as geometrically distributed with the same distributions as the age of information above.
At each successful acknowledgment, we can compute a no-change posterior using the latest acknowledgment age $D_n^s$ at the sensor.

We note that the receiver performs a test on the age of information \eqref{eq:ageofinfo} at each packet receipt $\gamma_k=1$ while the transmitter performs a test on the age of acknowledgment \eqref{eq:ageofack} at each acknowledgment receipt $\gamma_k^a=1$.
We note that the receiver will operate the detector more often than the transmitter does, as for an acknowledgment to be transmitted the packet needs to be received.

\section{SIMULATION}
\label{sec:simulation}
We now illustrate our results with a short simulation, and compare to a basic detector.
Additionally, we extend our simulation by considering that the eavesdropper's channel quality is unknown but bounded.

\begin{figure}
  \centering
  \includegraphics[width=0.45\textwidth]{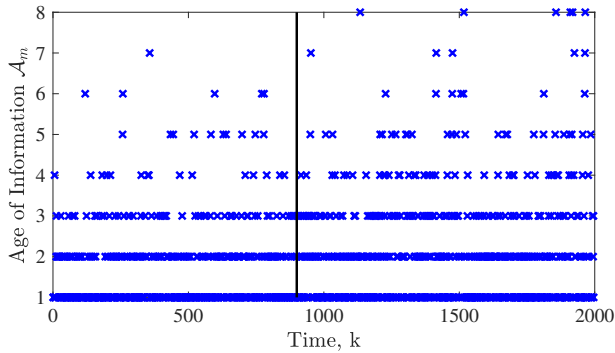}
  \caption{Age of information at the receiver, $\mathcal{A}_m$. The intruder begins blocking at $k=900$ marked as the vertical black line. The age of the measurements becomes slightly higher after the intrusion.}
  \label{fig:agedata}
\end{figure}

We consider a scalar dynamical system as proposed in Section~\ref{sec:remotestateest} with $A = 1.001$, $C=1$, $Q = 0.001$, and $R = 0.1$.
We choose the channel qualities as $\alpha=0.7$ for the legitimate receiver and $\alpha_a=0.9$ for the acknowledgment channel.
We consider that the eavesdropper's channel is perfect, remaining in sync until process time $900$ when the eavesdropper's channel quality changes to $\alpha_e = 0.8$ triggering the attacker to selectively block acknowledgments.
Following Section~\ref{subsec:qcd} the age of information is geometrically distributed with $\rho_1 = 0.63$ and $\rho_2 = 0.567$.
In MATLAB we simulate the process from $x_0 = 0.1$ for $2000$ steps, and the transmission receipts and acknowledgments using the uniform random generator.
Figure~\ref{fig:agedata} shows the age of the information at the sensor with intrusion time marked in vertical black.

\subsection{Comparison Detection Test}
\begin{figure}
  \centering
  \includegraphics[width=0.45\textwidth]{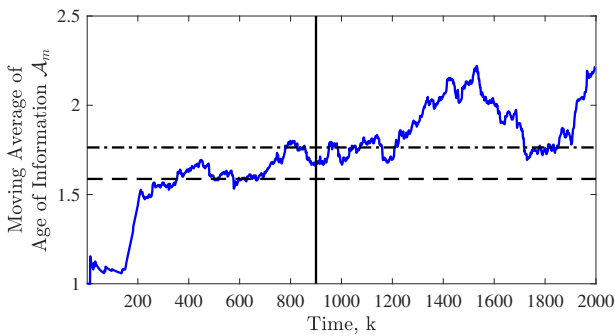}
  \caption{Moving average of the age of information over the last $150$ packet receipts at the legitimate estimator. Intrusion time marked with vertical solid line. Theoretical mean age under the geometric prior for no-intruder is $1/\rho_1 = 1.5873$ (marked in dashed black) and in the presence of the intruder is $1/\rho_2 = 1.7637$ (marked in dot-dashed black).}
  \label{fig:movingaverage}
\end{figure}

As a comparison we consider the average age of information and compare against the theoretical mean under the geometric distribution.
The mean age of information for no-intruder is $1/\rho_1 = 1.5873$ and in the presence of the intruder is $1/\rho_2 = 1.7637$.
We propose a moving average test over a window of length $N=150$ packet receipts.
Figure~\ref{fig:movingaverage} shows the moving average computed on the age of the information at the receiver in blue, with the intrusion marked with vertical black line, and the theoretical means in horizontal black (dashed for no-intruder, dot-dashed for intruder).
While it is visually apparent that the average age increases, it is challenging to quickly detect the presence of the intruder.

\subsection{Bayesian QCD Performance}
We apply the proposed Bayesian QCD test at the legitimate estimator.
The probability of change is chosen as $\rho_i = 5 \times 10^{-6}$.
For a probability of false alarm of $0.4$ we choose the threshold as $h=0.9875$.
We then compute the no-change posterior using the efficient method described in Section~\ref{subsec:efficientcalc}.

\begin{figure}
  \centering
  \includegraphics[width=0.45\textwidth]{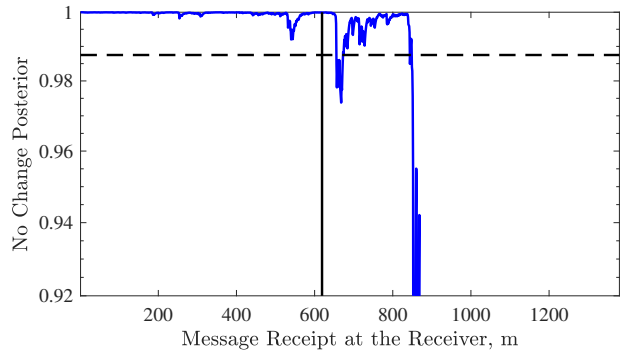}
  \caption{No change posterior QCD test statistic at the legitimate estimator against packet receipts $m$. The intrusion time ($k=900$, $m=619$) is marked in vertical black and the threshold for a probability of false alarm of $0.4$ is marked in horizontal black. The detection delay is $37$ packet receipts or $52$ time steps. The posterior drops very quickly after $m=850$ indicating strong probability of an intruder presence.}
  \label{fig:qcddetection}
\end{figure}

Figure~\ref{fig:qcddetection} shows the no-change posterior against the packet receipt $m$.
The intrusion time is marked in vertical black and the threshold is marked horizontally.
The receiver detects the intrusion at receipt $m=656$, equivalent to the real time step $k=952$, which is a detection delay of $37$ packet receipts and $52$ time steps.
We also observe that the posterior probability becomes very small after $m=850$, as the evidence that an intruder is in the acknowledgment channel builds.

\subsection{Mis-specified Bayesian QCD Performance}
In the above we have considered that the eavesdropper's channel quality is known exactly.
This would be unrealistic in a true scenario motivating us to investigate mis-specified scenarios.

We assume that the eavesdropper has an unreliable channel such that $P(\gamma_k^e = 1) = \alpha_e$ with $0 < \alpha_e < 1$.
The worst case would be when the eavesdropper's channel is closest to perfect or $\alpha_e = 1$.
Let us consider an upper bound on the set of $\tilde{\alpha}_e = 0.98$.
We now mis-specify the age of information in the post-change or intruder scenario to be geometrically distributed with parameter $\tilde{\rho}_2 = \alpha \alpha_a \tilde{\alpha}_e$ as follows $b_2(\mathcal{A}_m) \triangleq P(\mathcal{A}_m=\kappa)=\tilde{\rho}_2 (1 - \tilde{\rho}_2)^{\kappa-1}$ for $\kappa \ge 1$ and 0 otherwise.

This mis-specification might constitute the least favorable distribution of the uncertainty class that the legitimate estimator might expect of the eavesdropper's channel quality.
We note that using this distribution often has robustness properties in the \ac{QCD} framework, see for example the Minimax Robust result of \cite{Unnikrishnan2011IEEETransInfoTheoryMinimaxRobustQuickest}.
In this work, we do not formally show a robust result, we just explore the possibility of utilizing such a technique in simulation.

\begin{figure}
    \centering
    \includegraphics[width=0.45\textwidth]{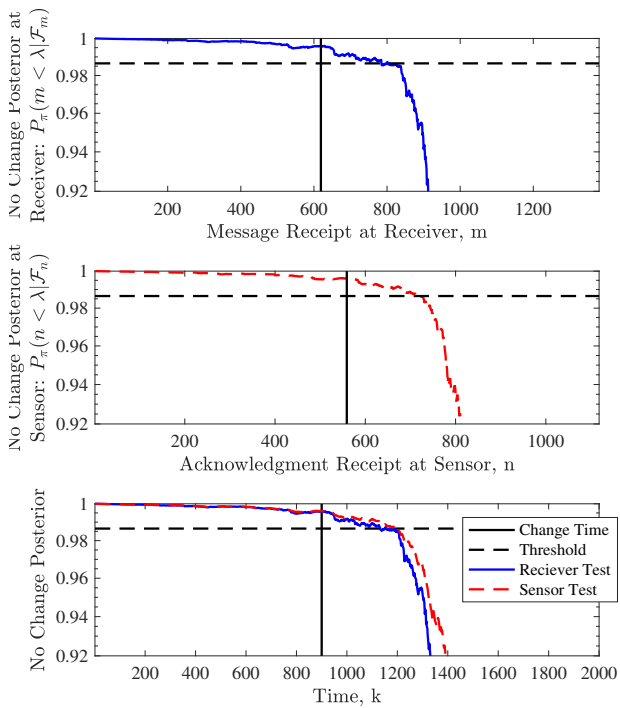}
    \caption{No change posterior QCD test statistic at the receiver against packet receipt (top) and sensor against acknowledgment receipt (middle), and together against process time $k$ (bottom). The threshold is chosen for probability of false alarm of $0.4$ (horizontal dashed black line). Intruder influences acknowledgments at $k=900$ (vertical solid black line). The posterior at the receiver (in blue) drops much slower than with exact knowledge as in Figure~\ref{fig:qcddetection} with increased detection delay of $234$ time steps. The detection delay at the sensor is $296$ time steps, slower than the legitimate estimator. This is expected as the sensor is using acknowledgments.}
    \label{fig:misspecCombined}
\end{figure}

To compare with our proposed rule we choose a threshold of $h=0.9865$ to achieve approximately the same false alarm rate of $0.4$ as above.
Figure~\ref{fig:misspecCombined} shows the no-change posterior at the receiver in blue against packet receipt $m$ (top) and against process time $k$ (bottom), with the mis-specified choice of channel quality for the eavesdropper.
The receiver detects the intrusion at receipt $m=785$, equivalent to the real time step $k=1134$, which is a detection delay of $166$ receipts and $234$ time steps.
The mis-specified stopping rule has larger detection delay than our Bayesian QCD rule with exact knowledge.
However, the mis-specified rule is designed with no knowledge of the eavesdropper's channel quality, just an assumption of the potential maximum channel quality.
The performance of the mis-specified rule is still greatly improved when compared to the comparison detector.

We also apply the mis-specified rule at the sensor with the same threshold.
Figure~\ref{fig:misspecCombined} shows the no change posterior at the sensor in dashed red against acknowledgment receipt $n$ (middle) and against process time $k$ (bottom).
The sensor detects the intrusion at acknowledgment receipt $n=722$, equivalent to the real time step $k=1196$, which is a detection delay of $163$ receipts and $296$ time steps.
The detection in process time $k$ is slower at the sensor than at the legitimate estimator.

Comparing receipt detection times is misleading as detection at the sensor was at $n=722$ compared to $m=785$ at the receiver.
However, a packet needs to be received before an acknowledgment is sent, so there have then been fewer acknowledgments than packets, and it is necessary to compare in process time $k$.
As expected, detection at the sensor is slower than at the legitimate estimator.

\section{CONCLUSIONS}
\label{sec:conclusion}
In this paper we propose a method to detect a stealthy eavesdropper in a remote state estimation process with encoded transmissions.
The eavesdropper selectively blocks packet receipt acknowledgments, from the legitimate receiver to the transmitter, to obtain a reliable state estimate, defeating the encoding scheme.
We illustrate that by utilizing statistical techniques such as Bayesian \ac{QCD}, it is possible to quickly detect the presence of an intruder in the acknowledgment channel.
This motivates future work in incorporating detection schemes to combat stealthy intruders, and designing transmission encoding with limited or unavailable packet receipt acknowledgment.

\addtolength{\textheight}{-12cm}   


\bibliographystyle{IEEEtran}
\bibliography{IEEEabrv,CybersecurityRefs}

\begin{thebibliography}{10}
\providecommand{\url}[1]{#1}
\csname url@samestyle\endcsname
\providecommand{\newblock}{\relax}
\providecommand{\bibinfo}[2]{#2}
\providecommand{\BIBentrySTDinterwordspacing}{\spaceskip=0pt\relax}
\providecommand{\BIBentryALTinterwordstretchfactor}{4}
\providecommand{\BIBentryALTinterwordspacing}{\spaceskip=\fontdimen2\font plus
\BIBentryALTinterwordstretchfactor\fontdimen3\font minus
  \fontdimen4\font\relax}
\providecommand{\BIBforeignlanguage}[2]{{%
\expandafter\ifx\csname l@#1\endcsname\relax
\typeout{** WARNING: IEEEtran.bst: No hyphenation pattern has been}%
\typeout{** loaded for the language `#1'. Using the pattern for}%
\typeout{** the default language instead.}%
\else
\language=\csname l@#1\endcsname
\fi
#2}}
\providecommand{\BIBdecl}{\relax}
\BIBdecl

\bibitem{Ishii2022SecurityResilienceControl}
H.~Ishii and Q.~Zhu, Eds., \emph{Security and Resilience of Control
  Systems}.\hskip 1em plus 0.5em minus 0.4em\relax Springer International
  Publishing, 2022.

\bibitem{Lucke2022IEEETransactionsonAutomaticControlCodingsecrecyremote}
M.~Lucke, J.~Lu, and D.~E. Quevedo, ``Coding for secrecy in remote state
  estimation with an adversary,'' \emph{{IEEE} Transactions on Automatic
  Control}, vol.~67, no.~9, pp. 4955--4962, 2022.

\bibitem{Tsiamis2020IEEETransactionsonAutomaticControlStateSecrecyCodes}
A.~Tsiamis, K.~Gatsis, and G.~J. Pappas, ``State-secrecy codes for networked
  linear systems,'' \emph{{IEEE} Transactions on Automatic Control}, vol.~65,
  no.~5, pp. 2001--2015, 2020.

\bibitem{Leong2019IEEETransactionsonAutomaticControlTransmissionSchedulingRemote}
A.~S. Leong, D.~E. Quevedo, D.~Dolz, and S.~Dey, ``Transmission scheduling for
  remote state estimation over packet dropping links in the presence of an
  eavesdropper,'' \emph{{IEEE} Transactions on Automatic Control}, vol.~64,
  no.~9, pp. 3732--3739, 2019.

\bibitem{Lu2021IEEETransactionsonAutomaticControlStealthyhackingsecrecy}
J.~Lu, D.~E. Quevedo, V.~Gupta, and S.~Dey, ``Stealthy hacking and secrecy of
  controlled state estimation systems with random dropouts,'' \emph{{IEEE}
  Transactions on Automatic Control}, 2021, {Early Access}.

\bibitem{Ding2019IEEETransactionsonControlofNetworkSystemsDoSAttacksRemote}
K.~Ding, X.~Ren, D.~E. Quevedo, S.~Dey, and L.~Shi, ``{DoS} attacks on remote
  state estimation with asymmetric information,'' \emph{{IEEE} Transactions on
  Control of Network Systems}, vol.~6, no.~2, pp. 653--666, 2019.

\bibitem{Ding2021IEEETransactionsonAutomaticControlRemoteStateEstimation}
K.~Ding, X.~Ren, A.~S. Leong, D.~E. Quevedo, and L.~Shi, ``Remote state
  estimation in the presence of an active eavesdropper,'' \emph{{IEEE}
  Transactions on Automatic Control}, vol.~66, no.~1, pp. 229--244, 2021.

\bibitem{Cheng2020IEEETransactionsonAutomaticControlEventBasedStealthy}
P.~Cheng, Z.~Yang, J.~Chen, Y.~Qi, and L.~Shi, ``An event-based stealthy attack
  on remote state estimation,'' \emph{{IEEE} Transactions on Automatic
  Control}, vol.~65, no.~10, pp. 4348--4355, 2020.

\bibitem{Zhang2018IEEETransactionsonControlofNetworkSystemsDoSAttackEnergy}
H.~Zhang, Y.~Qi, J.~Wu, L.~Fu, and L.~He, ``{DoS} attack energy management
  against remote state estimation,'' \emph{{IEEE} Transactions on Control of
  Network Systems}, vol.~5, no.~1, pp. 383--394, 2018.

\bibitem{Yang2022AutomaticaAdaptivedistributedKalman}
J.~Yang, W.-A. Zhang, and F.~Guo, ``Adaptive distributed {Kalman}-like filter
  for power system with cyber attacks,'' \emph{Automatica}, vol. 137, p.
  110091, 2022.

\bibitem{Naha2022IEEETransactionsonAutomaticControlSequentialdetectionReplay}
A.~Naha, A.~M.~H. Teixeira, A.~Ahlen, and S.~Dey, ``Sequential detection of
  replay attacks,'' \emph{{IEEE} Transactions on Automatic Control}, 2022,
  {Early Access}.

\bibitem{Tsiamis2018StateSecrecyCodes}
A.~Tsiamis, K.~Gatsis, and G.~J. Pappas, ``State-secrecy codes for stable
  systems,'' in \emph{American Control Conference ({ACC})}, Milwaukee, WI, June
  2018.

\bibitem{Anderson1979OptimalFiltering}
B.~D.~O. Anderson and J.~B. Moore, \emph{{Optimal Filtering}}, ser.
  Prentice-Hall Information and System Sciences Series, T.~Kailath, Ed.\hskip
  1em plus 0.5em minus 0.4em\relax Englewood Cliffs, N.J., USA: Prentice-Hall
  Inc., 1979.

\bibitem{elliott2008hidden}
R.~J. Elliott, L.~Aggoun, and J.~B. Moore, \emph{Hidden Markov models:
  estimation and control}.\hskip 1em plus 0.5em minus 0.4em\relax Springer
  Science \& Business Media, 2008, vol.~29.

\bibitem{Shiryaev1963TheoryofProbability&amp$mathsemicolon$ItsApplicationsOptimumMethodsQuickest}
A.~N. Shiryaev, ``On optimum methods in quickest detection problems,''
  \emph{Theory of Probability {\&} Its Applications}, vol.~8, no.~1, pp.
  22--46, 1963.

\bibitem{Unnikrishnan2011IEEETransInfoTheoryMinimaxRobustQuickest}
J.~Unnikrishnan, V.~V. Veeravalli, and S.~P. Meyn, ``Minimax robust quickest
  change detection,'' \emph{{IEEE} Transactions on Information Theory},
  vol.~57, no.~3, pp. 1604--1614, 2011.

\bibitem{Ford2020AutomaticainformativenessmeasurementsShiryaevs}
J.~J. Ford, J.~James, and T.~L. Molloy, ``{On the informativeness of
  measurements in Shiryaev's Bayesian quickest change detection},''
  \emph{Automatica}, vol. 111, 2020.

\end{thebibliography}

\end{document}